\documentclass[runningheads]{llncs}
\usepackage{graphicx}
\usepackage{times}
\usepackage{epsfig}
\usepackage{graphicx}
\usepackage{amsmath}
\usepackage{amssymb}
\usepackage{colortbl}
\usepackage[misc]{ifsym}
\usepackage{soul}
\usepackage[ruled,vlined]{algorithm2e}
\begin{document}
\title{Modality agnostic intracranial aneurysm detection through supervised vascular surface classification}

\titlerunning{Modality agnostic intracranial aneurysm detection}
%


\author{\v{Z}iga Bizjak\inst{1} \and
Bo\v{s}tjan Likar\inst{1} \and
Franjo Pernu\v{s}\inst{1} \and \v{Z}iga \v{S}piclin\inst{1}}

\authorrunning{Bizjak et al.}
\institute{${}^1$University of Ljubljana, Faculty of Electrical Engineering\\Tr\v{z}a\v{s}ka cesta 25, 1000 Ljubljana, Slovenia\\ \email{ziga.bizjak@fe.uni-lj.si}\\
}
%

\maketitle            
\begin{abstract}
Intracranial aneurysms (IAs) are generally asymptomatic and thus often discovered incidentally on angiographic scans like 3D DSA, CTA and MRA. Skilled radiologists achieved a sensitivity of 88\% by means of visual detection, which seems inadequate considering that prevalence of IAs in general population is 3-5\%. Deep learning models trained and executed on angiographic scans seem best-suited for IA detection, however, reported performances across different modalities is currently insufficient for clinical application. This paper presents a novel modality agnostic method for detection of IAs. First the triangulated surfaces of vascular structures were roughly extracted from the angiograms. For IA detection purpose, the extracted surfaces were randomly parcellated into local patches and then a translation, rotation and scale invariant classifier based on deep neural network (DNN) was trained. Test stage proceeded by mimicking the surface extraction and parcellation at several random locations, then the trained DNN model was applied for classification, and the results aggregated into IA detection heatmaps across entire vascular surface. For training and validation the extracted contours were presented to skilled neurosurgeon, who marked the locations of IAs. The DNN was trained and tested using three-fold cross-validation based on 57 DSAs, 5 CTAs and 5 MRAs and showed a 98.6\% sensitivity at 0.2 false positive detections per image. Experimental results show that proposed approach not only significantly improved detection sensitivity and specificity compared to state-of-the-art intensity based methods, but is also modality agnostic and thus better suited for clinical application. 
\end{abstract}

\section{Introduction}

Intracranial aneurysms (IAs) are abnormal vessel wall dilatations in the cerebral vasculature and, according to a study that included $94,912$ subjects~\cite{vlak2011prevalence}, have a high 3.2\% prevalence in the general population. Most IAs are small and it is estimated that 50-80\% do not rupture during a person's lifetime. Still, rupture of IA is one of the most common causes of subarachnoid hemorrhage (SAH) \cite{van2007subarachnoid}, a condition with 50\% mortality rate~\cite{etminan2019worldwide}. For small IAs with diameter $<5$ mm the chances of rupture are below 1\%, but increase with ageing and potential IA growth, whereas rupture risk is generally higher for larger IAs. Early detection of IAs is thus necessary to open a window of opportunity to mitigate rupture risk and/or to determine the best time and type of treatment.



Current clinical practice is to search for IAs by visual inspection of 3D angiographic images like CTA, MRA and DSA. Such visual inspection is time consuming (10-15 minutes per case) and is prone to human error. Even skilled experts have a sensitivity of 88\% for small IAs on the CTA~\cite{yang_small_2017}. This is among the reasons why in recent years many researchers have worked extensively on computer-assisted IA detection.


\begin{figure}[!t]
	\begin{center}
		\includegraphics[width=4.5in]{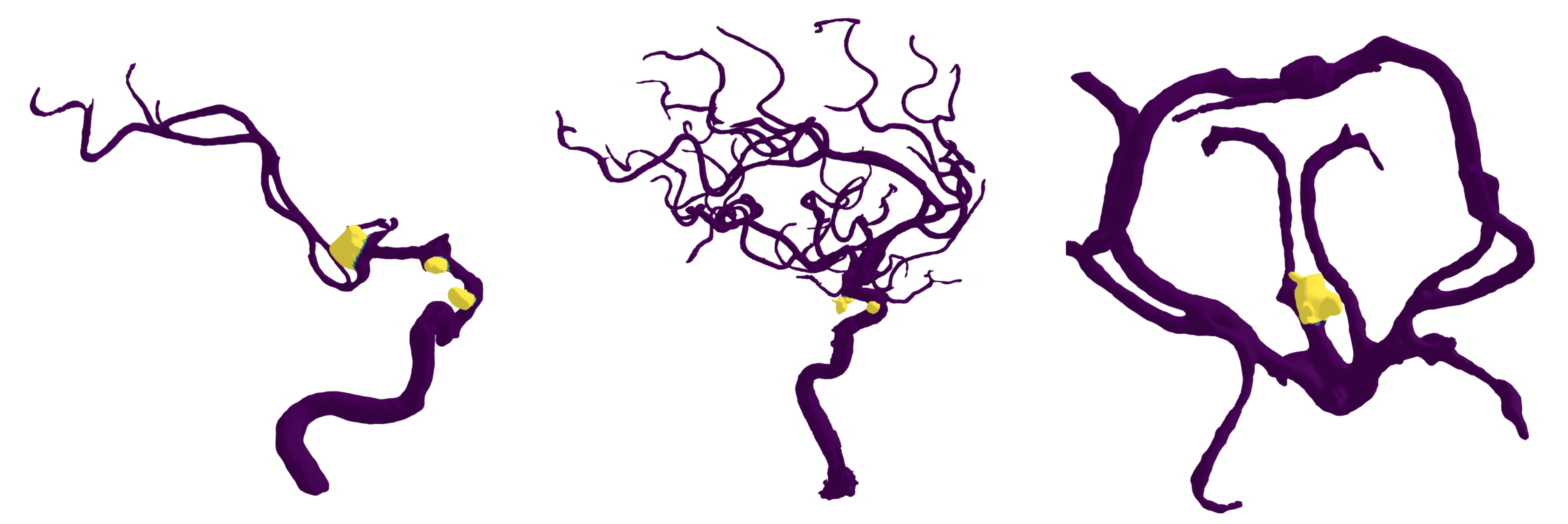}
	\end{center}
	
	\caption{\small Vascular 3D surface extracted from different modalities: MRA \textit{on the left}, DSA \textit{in the middle} and CTA \textit{on the right}. A skilled neurosurgeon marked the aneurysms, \textit{shown in yellow}.}
	\label{modalities}
\end{figure}

\subsection{Background}
Deep machine learning has become the most successful technique for IA detection. Nakao \textit{et al.}~\cite{nakao2018deep} utilized a 6-layer 3D convolutional neural network (CNN) to find IAs in TOF-MRA images. Similarly, Ueda \textit{et al.}~\cite{ueda2019deep} used ResNet-18, pre-trained to detect four vascular pathologies and then fine-tuned to detect IAs in TOF-MRA images. Sichtermann \textit{et al.}~\cite{sichtermann2019deep} employed the "Deepmedic" dual-pathway CNN with 11 layers and validated on 1.5T and 3T 3D TOF-MRA images. Jin \textit{et al.}~\cite{jin2019fully} trained and tested a combined U-net and BiConvLSTM on 2D DSA images. Duan \textit{et al.} \cite{duan2019automatic} proposed a two step algorithm: first stage involved a model to detect candidate regions on posterior communicating artery (PCoA) in the 2D DSA images, while second stage model was used to detect IAs. Dai~\textit{et al.}~\cite{dai2020deep} trained a CNN model to detect aneurysms on select 2D projections from 3D CTA images. Deep learning methods are capable of detecting IAs regardless of their shape and size, but require access to a massive annotated dataset. Another limitation is the use of intensity information, which renders these methods applicable for modalities they were trained on, while they also are not able to aggregate information from different modalities during training.

Modality-independent IA detection methods generally employ hand-crafted shape mapping and feature extraction. For instance, Jerman \textit{et al.}~\cite{jerman2015miccai} applied blobness and vesselness filters to 3D DSAs to locate potential IA locations and then encoded the filters' responses into rotation and scale invariant features using spherical harmonics based local neighborhood representation. The authors hypothesized that filters yield similar response on other angiographic modalities and that the trained model is transferable to other modalities, but this was not demonstrated. Another approach by the same authors~\cite{jerman2017aneurysm} requires vessel segmentation and applies intra-vascular distance mapping to the extracted vascular surface to cast unstructured 3D information into 2D gridded maps. Based on such maps, the CNNs were used for aneurysm site detection. The approach achieved a sensitivity of 100\% at 2.4 false positive IA detections per image. While results were promising, validation was limited with only 15 3D DSA images containing 21 IAs.
Zhou \textit{et al.}~\cite{zhou2019intracranial} cast IA detection as shape analysis. The segmented 3D cerebrovascular mesh model was parameterized into a planar flat-torus and local and global geometric features such as Gaussian curvature, shape diameter function and wave kernel signature were computed and input to three GoogleNet Inception V3 models. All three models were used to jointly detect IAs with adaptive learned weights. However, the authors selected for validation only bifurcations on the anterior cerebral artery (ACA) and internal carotid artery (ICA). A common disadvantage of the three mentioned methods is that hand-crafted routines may be suboptimal in terms of IA detection performance and are often computationally demanding.



A computer-assisted method that can detect IAs in all angiographic modalities is required, since 3DRA and DSA are used in interventional suites during acute states, while while CTA and MRA are used in population screening and prevention. Second, methods should aggregate information from different modalities during training because the amount of annotated data available for validation is generally limited due to ethical reasons. A big limitation of some studies is that they focused only on some areas of the cerebral vasculature (ICA, PCoA or ACA), regardless of the fact that IAs can develop elsewhere. For instance, 61\% of IAs are located on ICA and ACA while remaining 39\% of IAs are located in other areas of cerebral vasculature~\cite{brown2010unruptured}. Hence, it is not justified to focus only on certain locations as this may bias the detection results.


\subsection{Contributions}
In this paper we aim to detect aneurysms from 3D meshes obtained from CTA, MRA and DSA images as shown in Figure~\ref{modalities}. Unstructured point clouds were sampled from those meshes as input into deep neural network (DNN), which is trained to classify each point into IA or vessel. By aggregating the responses across locally sampled point clouds in test stage the accumulated heatmap was output as the basis for IA detection. To prove that our approach can be applied and transferred across different modalities, we trained our model on meshes obtained from one modality (DSA) and validated on meshes of other modalities (CTA, MRA). 

This paper has three major contributions: (1) a novel IA detection algorithm that is applicable to different angiographic modalities; (2) our approach achieved state-of-the-art detection sensitivity on the test and cross-modality validation datasets; (3) our approach achieved significantly lower false positive rate per image compared to current state-of-the-art methods.


\section{Data and Methods}

\subsection{Data acquisition}
Angiographic brain scans of 67 subjects were acquired at University medical center Ljubljana 
where the institutional ethics committee approved this study. For each subject included in the study informed patient consent was obtained.
Imaging followed a standard clinical imaging protocol and included 56 DSA, 5 CTA and 5 MRA angiographic scans of the cerebral vasculature. 
Sixty three subjects had one or more unruptured IAs (76 in total), while remaining 3 cases were aneurysm-free. The mean diameter of the observed IAs was $9.22$ mm, with 8\% small (diameter $<5$ mm), 66\% medium ($5$ mm $<$ diameter $<10$ mm) and 26\% large size (diameter $>10$ mm) aneurysms.


\subsection{Aneurysm detection}

\begin{figure}[!t]
	\begin{center}
		\includegraphics[width=3.5in]{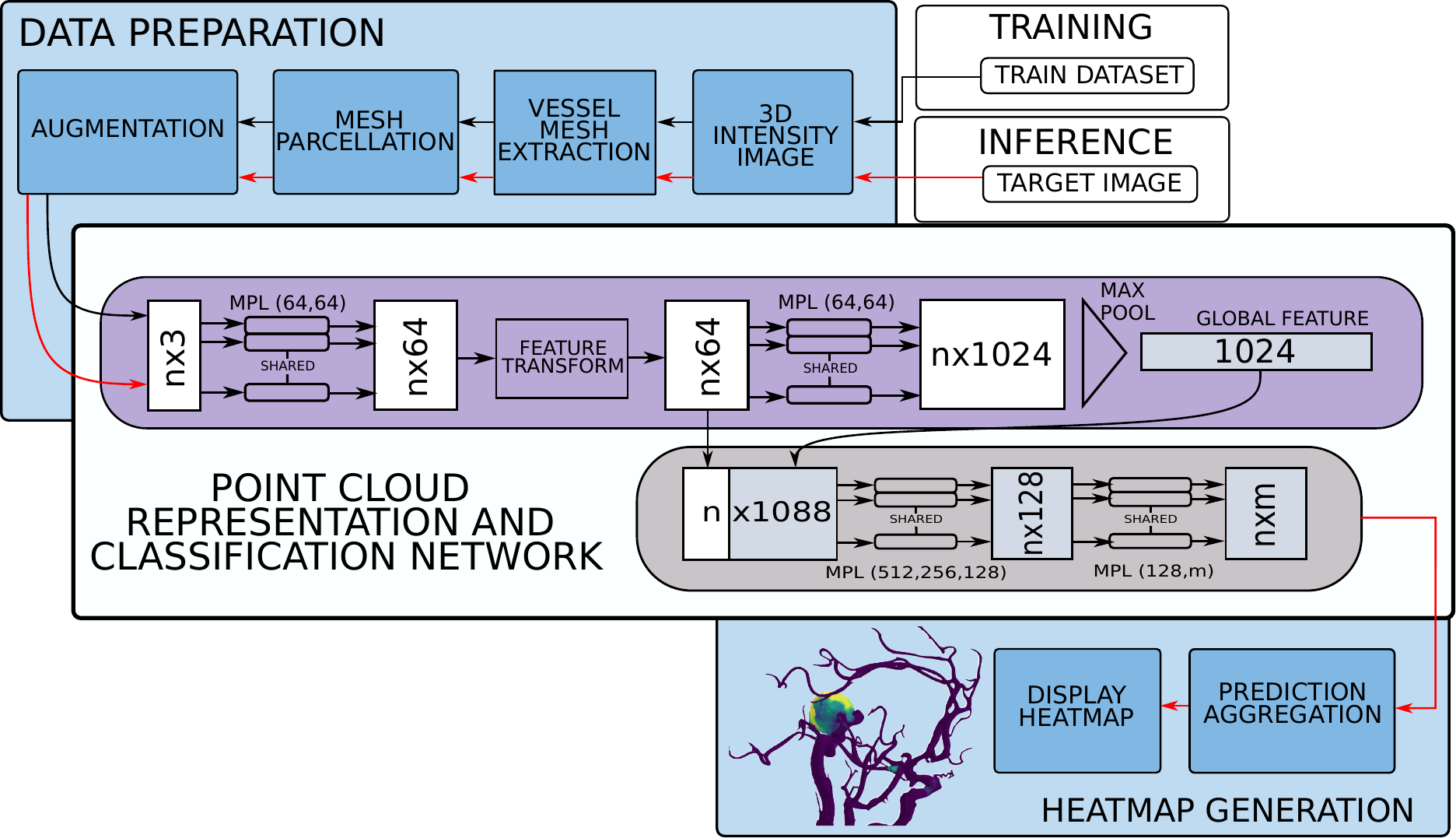}
	\end{center}
	\caption{\small Flowchart of the proposed aneurysm detection method.}
	\label{flowchart}
\end{figure}

This section presents a novel modality agnostic method for detection of IAs, which involves (1) triangulated surface mesh extraction from 3D angiographic images, (2) parcellation of the surface mesh into local patches of unstructured point clouds, (3) learning their invariant representation using DNN for classification of points into either an aneurysm or vessel and, in testing stage, (4) aggregation of predicted point classification obtained on local surface patches into IA detection heatmaps across entire vascular surface. For training and validating the IA detection method the extracted surface meshes were presented to skilled neurosurgeon, who annotated each IA by painting its surface. The flowchart of the proposed method is shown in Figure~\ref{flowchart}. The four steps are detailed in next subsections.

\subsubsection{(1) Surface mesh extraction.}
Cerebrovascular angiographic modalities are 3D images that depict vascular structures with high intensity. Other anatomical structures may also be depicted such as cranial bones in CTA and soft brain tissue in MRA. From the CTAs the cranial bone was removed using simple image cropping. Then we used interactive thresholding followed by the application of marching cubes and smooth non-shrinking algorithms \cite{larrabide2011three,cebral2001medical}. Resulting meshes were input to the aneurysm detection method.

\begin{figure}[!t]
	\begin{center}
		\includegraphics[width=4.7in]{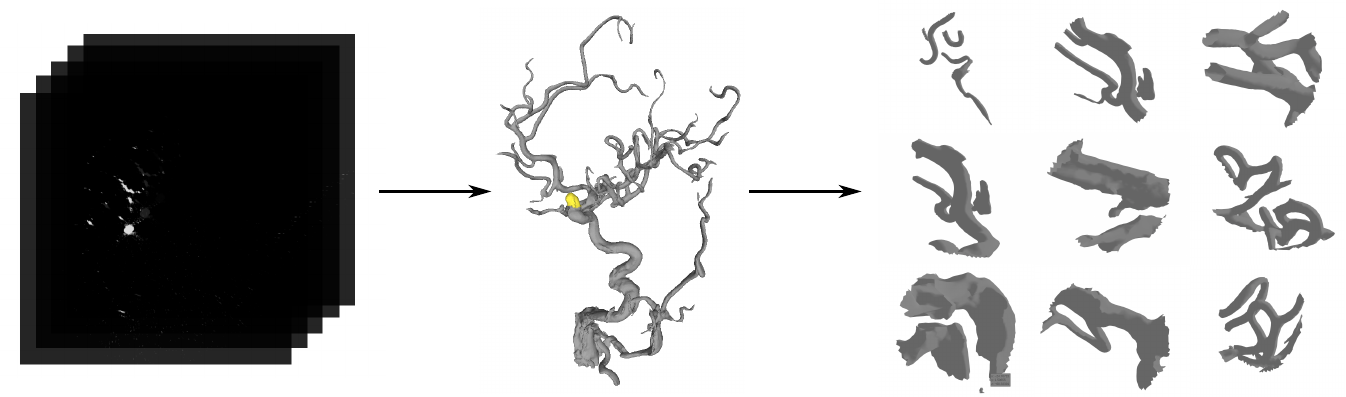}
	\end{center}
	\caption{\small Cerebral angiogram (\textit{left}) was interactively thresholded and marching cubes and smooth non-shrinking applied to extract a surface mesh (\textit{middle}). Yellow sections on the extracted mesh is the aneurysm annotated by skilled neurosurgeon. The obtained mesh was further used to extract unstructured point clouds (\textit{right}), each containing 3000 points.}
	\label{heatmap}
\end{figure}

\subsubsection{(2) Surface mesh parcellation.} Surface mesh was transformed into unstructured point cloud by randomly selecting seed points on the whole 3D vascular surface mesh and, for each seed, sampling 3000 closest mesh vertices according to geodesic distance. Hence, each point cloud contained exactly $N=3000$ points. For prediction purposes, the surface was repeatedly parcellated, with random seed selection, until every point on 3D vascular mesh was sampled at least once. For training purposes, each 3D mesh was parcellated into 170 point clouds, with ratio 50:120 of clouds containing aneurysm and vessel classes, respectively. 

\subsubsection{(3) Point cloud representation learning for classification.} We employ the PointNet  architecture~\cite{qi2017pointnet} to learn ordering, scale and rotation invariant representation of the input point clouds $\{x_i; i=1,\ldots,N\}, x_i\in\mathbb{R}^n$. The idea is to approximate a general function $f(\{x_1,\ldots,x_n\})\approx g(h(x_1),\ldots,h(h(x_n)))$ by finding two mapping functions, i.e. $h: \mathbb{R}^N \rightarrow \mathbb{R}^K$ and $g: \mathbb{R}^K \times \mathbb{R}^K \times \mathbb{R}^K\} \rightarrow \mathbb{R}$. Functions $h(\cdot)$ and $g(\cdot)$ are modeled as DNNs that take $N$ points as input, apply input and feature transformations, and then aggregate point features by max pooling. In this way, input shape is summarized by a sparse set of key points $K$. The network outputs classification scores for the two aneurysm and vessel classes. For DNN training we input the point clouds and manual annotation as output, then use negative log likelihood loss with Adam optimizer run for 100 epochs, learning rate of 0.001, decay rate of 0.5 and decay step 20. 

\begin{figure}
	\begin{center}
		\includegraphics[width=4.5in]{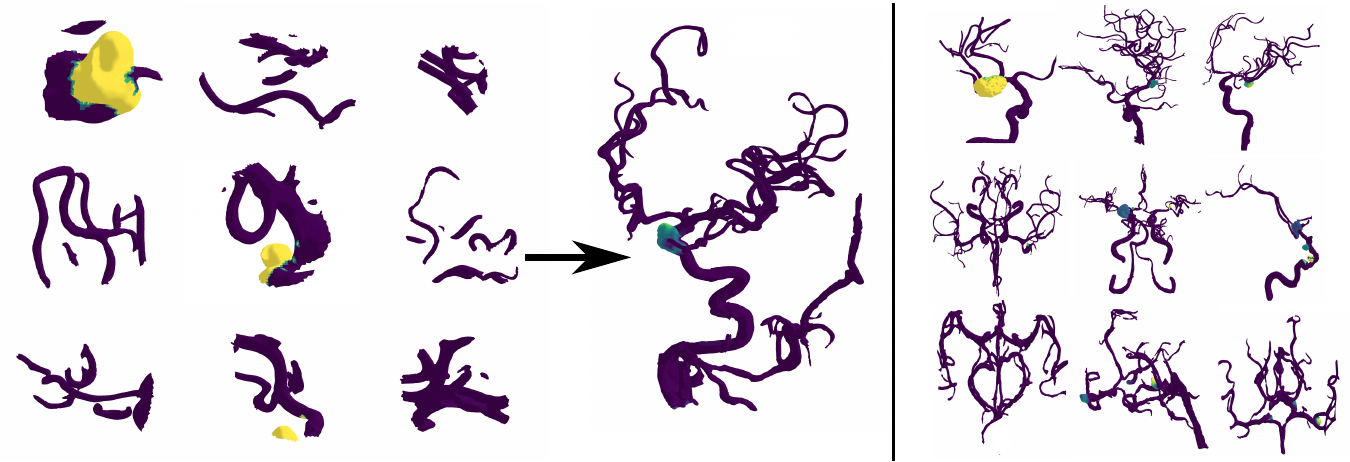}
	\end{center}
	
	\caption{\small Predictions on local point clouds were aggregated into aneurysm heatmap and superimposed onto 3D mesh (\textit{left}). Examplar predictions (\textit{right}) on DSA, MRA and CTA (\textit{top to bottom}).}
	\label{heat}
\end{figure}

\subsubsection{(4) Prediction aggregration.}
By extracting the surface mesh from the input image and parcellating into unstructured point clouds, the trained DNN was used to predict IA point labels for each of the point clouds. Then, the obtained soft class predictions were aggregated across all point clouds and their values normalized based on the number of point predictions. The final output was a heatmap for the aneurysm class as in Figure \ref{heat}, which indicates potential IA locations.

\subsubsection{Implementation.} 
Methods were implemented in Python 3.6 and PyTorch 1.4.0, and executed on a workstation with Intel i7 CPU, 32 GB RAM and NVidia GPU. 

\section{Experiments \& Results}

Performance of proposed aneurysm detection method was evaluated by comparing the obtained heatmaps to the manual annotation of IAs on the same surface as made by the skilled neurosurgeon. Simple threshold was applied to the heatmap to get binary results, then all surface sections larger than 50 connected points were labeled. If a labeled section overlapped with the manually annotated section this was considered a true positive (TP); if not, it was considered as false positive (FP). A false negative (FN) was noted in case no labeled section overlapped with the manual annotation. For threshold-free assessment we used free receiver operating characteristic curve (FROC), with the number of FPs per image (FP/I) on the horizontal axis and TP rate (sensitivity; TPR=TP/(TP+FN)) on the vertical axis, to evaluate detection performance. Area under FROC curve was also computed.

\subsubsection{Same-modality detection performance.}
The 57 DSA scans were applied for method training and evaluation based on three-fold cross-validation (3$\times$19). In each fold model was trained on 38 and tested on 19  images. On the test set the model successfully detected 63 out of 64 aneurysms (across all folds), while detecting only 13 FPs (0.2 FP/I). The sensitivity with respect to FP/I, as shown in FROC in Figure~\ref{FROC}, \textit{left}, was higher than 90\% with as low as 0.02 FP/I. The AUC was 0.91. 

\subsubsection{Cross-modality detection performance.}
\begin{figure}[!t]
	\begin{center}
		\includegraphics[width=4in]{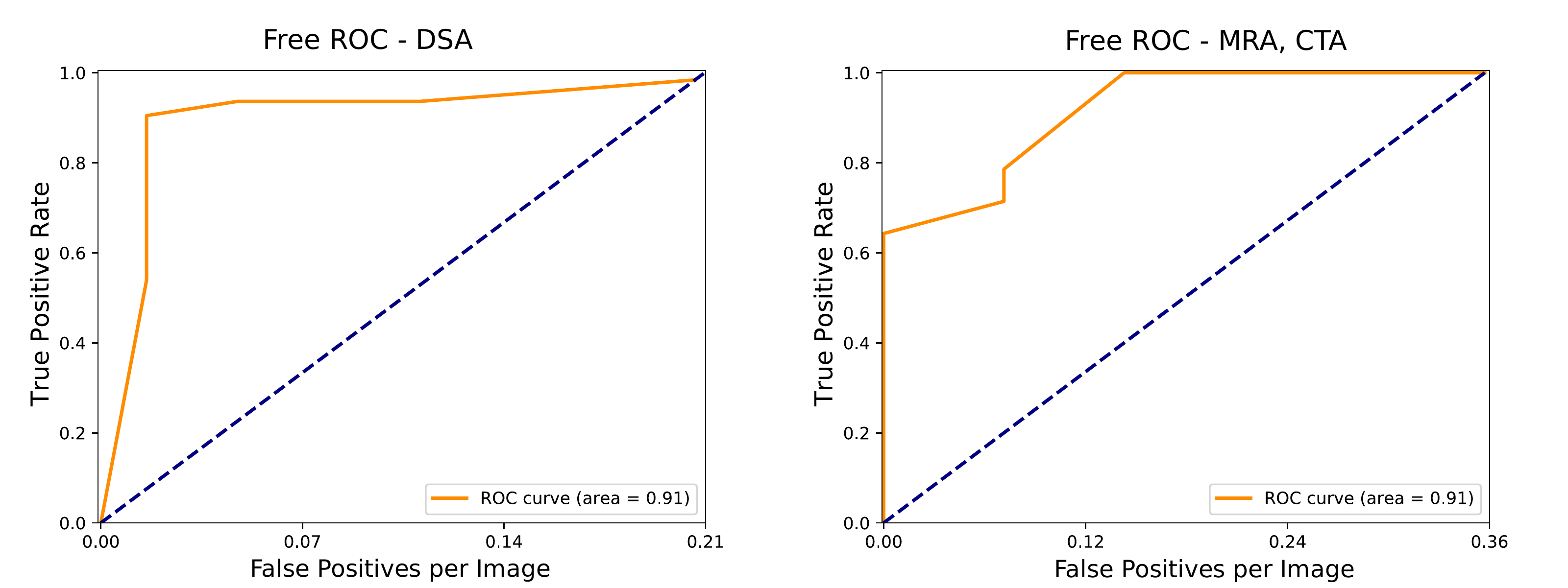}
	\end{center}
	\caption{\small Free receiver operating characteristic curve (FROC) across all modalities (\textit{left}) and for the cross-modality experiment (\textit{right}).}
	\label{FROC}
\end{figure}

Cross-modality validation set included surface meshes extracted from CTA and MRA images. These were used to validate the cross-modality performance using models trained only on the DSAs. The  method detected 12 out of 12 aneurysms, with 5 FPs (0.5 FP/I). The corresponding FROC is shown in Figure~\ref{FROC}, \textit{right} indicates 100\% sensitivity at 0.15 FP/I, and AUC of 0.91. These results are consistent with the performance of the method in case of same-modality detection.

\begin{table}[!t]
    \caption{Comparison of state-of-the-art and the proposed methods.}
	\begin{tabular}{|l|c|c|c|c|}
		\hline
		\textbf{Method}      & \textbf{Input Modality}                                                       & \textbf{Num. cases} & \textbf{FPs/image} & \textbf{TPR (\%)} \\ \hline\hline
		Dai et al. (2020) \cite{dai2020deep} & 2D CTA                              & 208            & 8.6                   & 91.8               \\ \hline
		Zhou et al. (2019)  \cite{zhou2019intracranial}      & \begin{tabular}[c]{@{}c@{}}mesh extracted from 3DRA,\\flat torus map with features \end{tabular}   & 121           & 0.8                      & 94.8             \\ \hline
		Sichtermann et al. (2019) \cite{sichtermann2019deep} & 3D TOF-MRA                                & 85            & 8.14                     & 87               \\ \hline
		Ueda et al. (2019)    \cite{ueda2019deep}   & 3D TOF-MRA                                        & 748           & 10                       & 91               \\ \hline
		Jin et al. (2019)   \cite{jin2019fully}      & 2D DSA                                           & 493           & 3.77                     & 89.3             \\ \hline
		Nakao et al. (2018)   \cite{nakao2018deep}     & 3D TOF-MRA                                      & 450           & 2.9                      & 94.2             \\ \hline
		Jerman et al. (2017)   \cite{jerman2017aneurysm}    & 3D DSA                                     & 15            & 2.4                      & \bf 100              \\ \hline

		Proposed & \begin{tabular}[c]{@{}c@{}}mesh extracted from\\ 3D DSA, CTA and MRA \end{tabular} & 67            & \textbf{0.2}                      & 98.6              \\ \hline
	\end{tabular}
	\label{table}
\end{table}

\subsubsection{Comparison to state-of-the-art.}
Across DSA (using cross-validation), CTA and MRA images the proposed method achieved an overall  sensitivity of 98.6\% (75/76 aneurysms detected). Method execution time was less than one minute per image. Table~\ref{table} summarizes the results of state-of-the-art and the proposed methods. Comparing our approach to state-of-the-art methods showed that ours achieved the lowest number of FP/I among all methods, while, at the same time, a higher or comparable level of sensitivity. 

The CNN-based method by Jerman \textit{et al.}~\cite{jerman2017aneurysm} achieved a slightly higher sensitivity compared to the proposed method, however, their database was rather small with only 15 3D DSA images and their method had substantially more false positives (2.4 FP/I). While their method could in principle accommodate multi-modal detection, its performance in such scenario was not verified. 

\section{Discussion \& Conclusion}

A novel method to detect aneurysms from 3D meshes obtained from CTA, MRA and DSA images was proposed. Method applied 3D mesh parcellation into unstructured point clouds, DNN for point classification and aggregation of predictions into the original 3D mesh. Since it is independent of intensity, the method is applicable to different angiographic modalities. The evaluated performance surpassed that of state-of-the-art detection method (Table~\ref{table}).  Namely, the proposed method achieved high sensitivity on the same- (98.6\%) and cross-modality (100.0\%) test datasets, with the least false positives ($<0.2$ per image) compared to all other methods. More importantly, training was executed only with DSA cases, whilst the sensitivity (TPR) and specificity (FP/I) were consistent when testing the method on DSA or CTA and MRA.

To the best of our knowledge, our method was the only one tested in the most difficult multi-modal scenario, i.e. training on one angiographic modality and testing on others. In our case, we used DSA scans for training and tested on CTA and MRA scans. Note that CTA and MRA scans had much lower resolution ($<$1/3 the resolution of DSA) and additional artifacts such as overlapping bony and tissue structures that adversely impact segmentation quality. But still, the observed IA detection performance was excellent and consistent with the performance in case of same-modality detection.

One limitation of this study is the number of angiographic scans used for validation. With increasing number of scans, from different modalities, more challenging cases, such as small aneurysms, would inevitably arise. Our dataset contained 7 small aneurysms (diameter $<$5 mm). The benefit of our approach, however, is that it accommodates training datasets aggregated across modalities and, albeit not tested here due to limited cases, we expect this would prove beneficial to the method performance.



Since our method was applied and validated to reliably detect intracranial aneurysms on three most common cerebrovascular angiographic modalities (CTA, DSA and MRA) it seems suitable for application in computer-assisted detection systems. The output prediction in form of the 3D heatmap superimposed on the extracted vascular surface can be used to help neurosurgeon detect aneurysms. Currently the visual inspection and detection may take from 5 to 15 minutes per image, depending on the image modality and the position and size of the aneurysm. Because the search for aneurysms is still done manually, the tool to visualize 3D vascular system and propose possible aneurysm locations would render (small) aneurysm detection more sensitive (visual sensitivity on CTAs was 88\%~\cite{yang_small_2017}) and more reliable. By using our method it take less then a minute to detect  and visualize aneurysm locations, saving a substantial amount of manual effort and reduces inspection time.

\bibliographystyle{splncs04}
\bibliography{mybibliography}

\end{document}